\documentclass[aps,prd,showpacs,preprintnumbers]{revtex4}
\usepackage{graphicx}
\usepackage{epsfig}
\usepackage{amsmath}
\usepackage{epsf,amssymb,amsfonts,amsmath}

\newcommand{\beqn}{\begin{eqnarray}}
\newcommand{\eeqn}{\end{eqnarray}}

\newcommand{\be}{\begin{equation}}

\newcommand{\ee}{\end{equation}}
\newcommand{\bea}{\begin{eqnarray}}
\newcommand{\eea}{\end{eqnarray}}

\newcommand{\ep}{\epsilon}
\usepackage{bm}

\begin{document}
\title{Survival of charged $\rho$ condensation at high temperature and density}

\author{Hao Liu$^{1}$}
\email{haoliu@mail.ihep.ac.cn}
\author{Lang Yu$^{1}$}
\email{yulang@mail.ihep.ac.cn}
\author{Mei Huang$^{1,2}$}
\email{huangm@mail.ihep.ac.cn}
\affiliation{$^1$ Institute of High Energy Physics, Chinese Academy of Sciences,
Beijing 100049, China}
\affiliation{$^2$ Theoretical Physics Center for Science Facilities,
Chinese Academy of Sciences,
Beijing 100049, China}

\begin{abstract}
The charged vector $\rho$ mesons in the presence of external magnetic fields at  finite  temperature $T$ and chemical potential $\mu$ have been investigated in the framework of the Nambu--Jona-Lasinio model. We compute the masses of charged $\rho$ mesons  numerically as a function of the magnetic field for different  values of  temperature and chemical potential. The self-energy of the $\rho$ meson contains the quark-loop contribution, i.e. the leading order contribution in $1/N_c$ expansion. The charged $\rho$ meson mass decreases with the magnetic field  and drops to zero at a critical magnetic field $eB_c$, which means that the charged vector meson condensation, i.e. the electromagnetic superconductor can be induced  above the critical magnetic field. Surprisingly, it is found that the charged $\rho$ condensation can even survive at high temperature and density. At zero temperature, the critical magnetic field just increases  slightly with the chemical potential, which indicates that the charged $\rho$ condensation might occur inside compact stars. At zero density,  in the temperature range $0.2-0.5~ {\rm GeV}$, the critical magnetic field for charged $\rho$ condensation is in the range of $0.2-0.6~ {\rm GeV}^2$, which indicates that the high temperature electromagnetic superconductor could be created at LHC.
\end{abstract}
\pacs{13.40.-f, 25.75.-q, 11.10.Wx }
\maketitle
\section{Introduction}
Quantum chromodynamics(QCD) is widely accepted as the fundamental theory for strong interaction. Exploring QCD phase structure  and properties of QCD matter at extreme conditions is one of the most important topics of high energy nuclear physics.  Recently, it becomes more and more attractive to study properties of QCD matter under strong magnetic fields. Strong magnetic fields could exist in various physical systems, e.g. the strength of magnetic fields can reach up to $10^{14}$~G in magnetars, and the magnitude of $10^{18-23}$ G of  magnetic fields may appear in early universe. Especially, in the laboratory, strong magnetic fields with the strength of $10^{18-20}$~G (amounts to $ eB\sim0.1-1$~GeV$^2$) can be generated in the early stage of  non-central heavy ion collisions \cite{Skokov:2009qp,Deng:2012pc} at the Relativistic Heavy Ion Collider (RHIC) or the LHC. Therefore, heavy ion collisions offer a unique platform to probe properties of QCD vacuum and hot/dense quark matter under strong magnetic fields.

Magnetic fields can modify the spectrum of charged particles, the energy levels of free charged particles in an external magnetic field parallel to $z$ axis are given as $E_{n,s_z}^2(p_z)=p_z^2+(2n-2sgn(q)s_z+1)|qB|+m^2$, where the nonnegative integer $n$ is the Landau level, $s_z$ is the spin projection, and $p_z$ is the particle's momentum in the direction of the magnetic field. Hence, $m^2_{\pi^{\pm}}(B)=m^2_{\pi^{\pm}}(B=0)+eB$ and $m^2_{\rho^{\pm}}(B)=m^2_{\rho^{\pm}}(B=0)-eB$ describe the minimal effective squared masses of charged $\pi$ and charged $\rho$, respectively. As the magnetic field increases, the charged $\pi$ mesons become heavier while the polarized charged $\rho$ mesons become lighter. Chernodub firstly proposed that charged $\rho$ meson can form a superconductor state when magnetic field is stronger than the critical magnetic field $eB_c\approx m^2_{\rho^{\pm}}(B=0)\approx0.6$ GeV$^2$ \cite{Chernodub:2010qx,Chernodub:2011mc}. This characterizes a quantum phase transition caused only by the magnetic field, which is a surprising result.

It is still under investigation whether there exists a vacuum superconductor. In Ref.\cite{Hidaka:2012mz}, it was argued that the vacuum
superconductor phase cannot appear due to the Vafa-Witten theorem. However, the authors in Ref.\cite{Li:2013aa} argued that the
Vafa-Witten theorem would not forbid the charged $\rho$ meson condensation in an external magnetic field. More studies on the
charged $\rho$ condensation have been done in different frameworks such as the Nambu-Jona-Lasinio (NJL) model \cite{Frasca:2013kka,Liu:2014uwa}, lattice QCD \cite{Hidaka:2012mz,Luschevskaya:2014mna}, the gauge/gravity correspondence \cite{Callebaut:2011uc,Ammon:2011je}, the relativistic Hamiltonian technique \cite{Andreichikov:2013zba}, and the Dyson-Schwinger equations \cite{Wangkunlun:2013}. On one hand,  it was observed that masses of charged $\rho$ mesons decrease firstly and then increase with the magnetic field, and cannot drop to zero, as shown by lattice calculations \cite{Hidaka:2012mz}, Dyson-Schwinger equations \cite{Wangkunlun:2013} and by solving the meson spectra in a relativistic quark-antiquark system with the relativistic Hamiltonian technique in Ref.\cite{Andreichikov:2013zba}. On the other hand, some other results showed that the charged $\rho$ masses indeed decrease with magnetic field and becomes massless at a critical magnetic field as shown in the NJL model \cite{Chernodub:2011mc,Frasca:2013kka,Liu:2014uwa} 
and in the holographic QCD model \cite{Callebaut:2011uc}.

Furthermore, for charged $\rho$ condensation, the obtained results for the critical magnetic field are quite different in different calculations.
For example, the polarized $\rho^{\pm}$ condensstion is expected to occur at the critical magnetic field $eB_c\approx 1$ GeV$^2$ in lattice QCD \cite{Luschevskaya:2014mna}, and $eB_c\approx 1.08m_{\rho}^2(B=0)$ is given in Ref.\cite{Callebaut:2011uc} in a holographic approach. Even in the framework of the NJL model, the results for the critical magnetic field differ a lot. The critical magnetic field $eB_c>1~{\rm GeV}^2$ was obtained in Ref.\cite{Chernodub:2011mc} by neglecting the quark-loop contribution, and $eB_c=0.98M_q^2$ was given in Ref.\cite{Frasca:2013kka}, where $M_q$ is the quark mass. In  Ref.\cite{Liu:2014uwa}, we have obtained the critical magnetic field $eB_c\approx 0.2~{\rm GeV}^2$ in the NJL model by calculating the $\rho$ meson
polarization function to the leading order of $1/N_c$ expansion.

By taking the quark propagator in the Ritus form as well as the Landau level representation, we have obtained the same results for the critical magnetic field $eB_c\approx 0.2~{\rm GeV}^2$, which is only 1/3 of the results from the point-particle calculation in Ref.\cite{Chernodub:2010qx}. Our results suggests that the charged $\rho$ condensation can be realized  in nature much easier. In this work, we are going to study the charged $\rho$ mesons in a magnetic field at finite temperature and chemical potential in the framework of the NJL model, and investigate the temperature and density region for the survival of charged $\rho$ condensation. This paper is organized as follows:  In Sec. II, we give a general description of the two-flavor NJL model including the effective four-quark interaction in the vector channel, and derive the vector meson mass under magnetic fields at finite temperature and chemical potential.
We give numerical results and analysis in Sec. III and finally in Sec. IV the discussion and conclusion is given.

\section{Model and formalism }
\label{Sec-Model}

\subsection{The two-flavor magnetized NJL model}
In this paper, we investigate the properties of the charged $\rho$ meson under the magnetic fields with nonzero temperature $T$ and chemical potential  $\mu$ in the framework  of the SU(2) NJL model. The NJL model, seen as a low-energy approximation of QCD, is regarded as an effective theory to study the properties of mesons ~\cite{Nambu:1961tp,Nambu:1961fr, Klimt:1989pm, Vogl:1991qt, Klevansky:1992qe,Hatsuda:1994pi}.
In our model, the Lagrangian density is given by
\begin{eqnarray}
{\cal{L}}&=&\bar{\psi}(i\not{\!\!D}-\hat{m}+\mu\gamma^0)\psi+G_{S}\left[(\bar{\psi}\psi)^2
   +(\bar{\psi}i\gamma^5\vec{\tau}\psi)^2\right] \nonumber \\
 & & -G_{V}\left[(\bar{\psi}\gamma^{\mu}\mathbf{\tau}^a\psi)^2+
  (\bar{\psi}\gamma^{\mu}\gamma^5\mathbf{\tau}^a\psi)^2 \right] -\frac{1}{4}F_{\mu\nu}F^{\mu\nu}.
\label{eq:L:basic}
\end{eqnarray}
In the above equation,  $\psi$ corresponds to the quark field of  two light  flavors $u$ and $d$, $\hat{m}=diag(m_u,m_d)$ is the
current quark mass matrix of $u$ and $d$ quarks, $\tau^a=(I,\vec{\tau})$ with $\vec{\tau}=(\tau^1,\tau^2,\tau^3)$ corresponding
to the isospin Pauli matrices, and $G_S$ and $G_V$ are the coupling constants with respect
to the scalar (pseudoscalar) and the vector (axial-vector) channels, respectively.
The covariant derivative, $D_{\mu}=\partial_{\mu}-i q_f A_{\mu}^{ext}$, couples quarks to an
external magnetic field $\bm{B}=(0,0,B)$ along the positive $z$ direction via a background
field, for example, $A_{\mu}^{ext}=(0,0,Bx,0)$. $q_f=(-1/3,2/3)$ is defined as the electric
charge of the quark field. The field strength tensor $F_{\mu\nu}$ is defined as usual by
$F_{\mu\nu}=\partial_{[\mu}A_{\nu]}^{ext}$, with $A_{\mu}^{ext}$ fixed as above.

The semibosonized Lagrangian which is equivalent to the above Lagrangian is given by
 \begin{eqnarray}\label{NE2b}
{\cal{L}}_{sb}&=&\bar{\psi}(x)\left(i\gamma^{\mu}D_{\mu}-\hat{m}+\mu\gamma^0\right)\psi(x)
-\bar{\psi}\left(\sigma+i\gamma_5\vec{\tau}\cdot\vec{\pi}\right)\psi\nonumber\\
&&-\frac{(\sigma^2+\vec{\pi}^2)}{4G_S}+\frac{(V_{\mu}^{a}V^{a\mu}+A_{\mu}^{a}A^{a\mu})}{4G_V}-\frac{B^2}{2},
\end{eqnarray}
where the Euler-Lagrange equation of motion for the auxiliary fields lead to the constraints
\begin{eqnarray}\label{NE3b}
\sigma(x)&=&-2G_S<\bar{\psi}(x)\psi(x)>,\\
\vec{\pi}(x)&=&-2G_S<\bar{\psi}(x)i\gamma_5\vec{\tau}\psi(x)>, \\
V_\mu^a(x)&=&-2G_V<\bar{\psi}(x)\gamma_\mu\tau^a \psi(x)>, \\
A_\mu^a(x)&=&-2G_V<\bar{\psi}(x)\gamma_\mu\gamma^5\tau^a \psi(x)>.
\end{eqnarray}
The quarks obtain the dynamical masses by the quark-antiquark condensation, so the constituent quark mass $M$ of  $u$ and $d$ is defined by
\be \label{eq:gapequationmq}
M=m_0-2G_S <\bar{\psi}\psi>,
\ee
where we  assume $m_u=m_d=m_0$.  
To determine the dynamical quark mass, in other words, the $\sigma$ condensation, we need to minimize the effective potential. The one-loop level effective potential of the model is given by
 \begin{eqnarray} \label{eq:effectivepotential}
 \Omega&=&\frac{\sigma^2}{4G_S}+\frac{B^2}{2}-3\sum_{q_f\in\{\frac{2}{3},-\frac{1}{3}\}}\frac{|q_feB|}{\beta}\sum_{p=0}^{+\infty}\alpha_p\int_{-\infty}^{+\infty}\frac{dp_3}{4\pi^2} \nonumber\\ &&\left\{\beta E_q+\ln \left(1+e^{-\beta(E_q+\mu)}\right)+\ln \left(1+e^{-\beta(E_q-\mu)}\right)\right\},\nonumber\\
 \end{eqnarray}
 where $\beta=\frac{1}{T}$ and $\alpha_p=2-\delta_{p0}$ is the spin degeneracy factor.

In the NJL model, mesons are constructed by the infinite sum of quark-loop chains by using random phase approximation. We calculate the $\rho$ meson polarization function to the leading order of $1/N_c$ expansion. The propagator of the $\rho$ meson $D_{ab}^{\mu\nu}(q^2)$ can be obtained from the one quark loop polarization $\Pi_{\mu\nu,ab}(q^2)$ via the Schwinger-Dyson equation and takes the form of
\begin{eqnarray}
\label{rhopropagator}
\left[-iD_{ab}^{\mu\nu}\right]&=&\left[-2iG_V\delta_{ab}g^{\mu\nu}\right] +\left[-2iG_V\delta_{ac}g^{\mu\lambda}\right]
     \left[-i\Pi_{\lambda\sigma,cd}\right] \left[-iD^{\sigma\nu}_{db}\right],
\end{eqnarray}
where a, b, c, d are isospin indices and $\mu$, $\nu$, $\lambda$, $\sigma$ are Lorentz indices. The one-loop polarization $\Pi^{\mu\nu,ab}(q^2)$  is given by
\begin{eqnarray}
\label{polarization}
\Pi^{\mu\nu,ab}(q^2)&=&-i\int d^4x~e^{i q\cdot x}Tr[\gamma^{\mu}\tau^a S_{Q}(x,0)\gamma^{\nu}\tau^b S_{Q}(0,x)].
\end{eqnarray}
Here,  $S_Q(x,y)$ is the Ritus fermion propagator \cite{Fayazbakhsh:2012vr,Fukushima:2009ft}
\be\label{Ritus}
S_{Q}(x,y)=i\sum_{p=0}^{\infty}\hspace{-0.5cm}\int{\cal{D}}\tilde{p}~e^{-i\tilde{p}\cdot
(x-y)}P_{p}(x_{1})D_{Q}^{-1}(\bar{p})~P_{p}(y_{1}),
\ee
arising from the solution of Dirac equation under a uniform magnetic field using Ritus eigenfunction method. In Eq.(\ref{Ritus}), $\tilde{p}=(p_{0},0,p_{2},p_{3})$, ${\cal{D}}\tilde{p}\equiv
\frac{dp_{0}dp_{2}dp_{3}}{(2\pi)^{3}}$, and $P_{p}(x_{1})$ is given by
\begin{equation}
\hspace{-0.8cm}P_{p}(x_{1})\equiv\frac{1}{2}[f_{p}^{+s}(x_{1})+\Pi_{p}f_{p}^{-s}(x_{1})]+\frac{is_Q}{2}[f_{p}^{+s}(x_{1})-\Pi_{p}f_{p}^{-s}(x_{1})]
\gamma^{1}\gamma^{2},
\end{equation}
where, $\Pi_{p}\equiv 1-\delta_{p,0}$ considering the spin degeneracy in the LLL, $s_Q\equiv sgn(Q eB)$. The functions $f_{p}^{\pm s}(x_{1})$ are defined by
\begin{eqnarray}
\begin{array}{rclcrcl}
f_{p}^{+s}(x_{1})&\equiv&\phi_{p}\left(x_{1}-s_{Q}p_{2}\ell_{B}^{2}\right),&&
p&=&0,1,2,\cdots,\nonumber\\
f_{p}^{-s}(x_{1})&\equiv&\phi_{p-1}\left(x_{1}-s_{Q}p_{2}\ell_{B}^{2}\right),&&
p&=&1,2,3,\cdots,
\end{array}
\hspace{-0.4cm}\nonumber\\
\end{eqnarray}
where $\phi_p(x)$ is a function of Hermite polynomials $H_p(x)$ in the form
\begin{eqnarray}
\phi_{p}(x)\equiv a_{p}\exp\left(-\frac{x^{2}}{2\ell_{B}^{2}}\right)H_{p}\left(\frac{x}{\ell_{B}}\right).
\end{eqnarray}
Here, $a_{p}\equiv (2^{p}p!\sqrt{\pi}\ell_{B})^{-1/2}$ is the
normalization factor and $\ell_{B}\equiv |QeB|^{-1/2}$ is the
magnetic length. In Eq.(\ref{Ritus}), $D_Q(\bar{p})=\gamma\cdot\bar{p}_Q-M$ with the Ritus four momentum $\bar{p}=(p_{0},0,-s_{Q}\sqrt{2|QeB|p},p_{3}).$ Note that  Q is a $2\times2$ matrix in the flavor space, so the functions $f_{p}^{\pm s}(x_{1})$ are matrices in the flavor space.

Using the propagator in Eq.(\ref{Ritus}), the one quark loop polarization function $\Pi_{\mu\nu,ab}(q^2)$ is given by
\begin{eqnarray}
\label{rituspolarization}
\Pi^{\mu\nu,ab}(q^2) & = & i\sum_{p,k=0}^{\infty}\int\ {\cal{D}}\tilde{p}\ {\cal{D}}\tilde{k}
                 \int d^{4}x  e^{-i(\tilde{p}-\tilde{k}-q)\cdot x}\Lambda^{\mu\nu,ab}_{pk}(\bar{p},\bar{k},x_1),
\end{eqnarray}
where
\begin{eqnarray}
\label{Lambda}
\Lambda^{\mu\nu,ab}_{pk}(\bar{p},\bar{k},x_1)&=&
Tr\big[\gamma^{\mu}\mathbf{\tau}^{a}P_{p}(x_{1})D^{-1}_{Q}(\bar{p})P_{p}(0)
 \gamma^{\nu}\mathbf{\tau}^{b}
 K_{k}(0)D^{-1}_{Q}(\bar{k})K_{k}(x_{1})\big].
\end{eqnarray}
 We can get the boundary conditions $q_i=p_i-k_i$ $( i=0,2,3)$ by integrating over $x_0, x_2, x_3$ and then over $p_0, p_2, p_3$ components in Eq.(\ref{rituspolarization}). So we get
\begin{eqnarray}
\label{rituspolarization1}
\Pi^{\mu\nu,ab}(q^2) & = &i\sum_{p,k=0}^{\infty}\int \frac{dk_0dk_3}{(2\pi)^3}\int dk_2dx_1 e^{iq_1x_1}\Lambda^{\mu\nu,ab}_{pk}(\bar{p},\bar{k},x_1)\big{|}_{b.c}.
\end{eqnarray}
In above equation Eq.(\ref{rituspolarization1}), the  $b.c$ is short for boundary conditions.

\subsection{Charged $\rho$ meson  in a magnetic field at nonzero $T$ and $\mu$}

In the rest frame of $\rho$, the Lorentz and flavor structure of the one quark loop polarization allows for the following decomposition
\begin{eqnarray}
\label{tensordecompose}
\Pi_{ab}^{\mu\nu}(q^2)&=&[\Pi_1^2(q^2) P_{1}^{\mu\nu}+\Pi_2^2(q^2) P_{2}^{\mu\nu}+\Pi_3^2(q^2) L^{\mu\nu}+\Pi_4^2(q^2)u^{\mu}u^{\nu}]\delta_{ab},
\end{eqnarray}
where $u^{\mu}=(1,0,0,0)$ is the 4-velocity  in the rest frame. Here, we define the spin projection operators
\bea
P_{1}^{\mu\nu} & = & -\epsilon_2^{\mu}\epsilon_2^{\nu*}, \, (s_z= -1\, \text{for} \, \rho), \\
P_{2}^{\mu\nu} & = & -\ep_1^{\mu}\ep_1^{\nu*}, \, (s_z= 1\, \text{for} \, \rho),\\
L^{\mu\nu} & = & -b^{\mu}b^{\nu}, \, (s_z=0 \, \text{for} \, \rho),
\eea
where $b^{\mu}=(0,0,0,1)$ corresponds to the external magnetic field direction and $\ep_1^{\mu}$, $\ep_2^{\mu} $ are  the right and left-handed polarization vectors
\bea
\ep_1^{\mu} &=& \frac{1}{\sqrt{2}}(0,1,i,0), \\
\ep_2^{\mu} &=& \frac{1}{\sqrt{2}}(0,1,-i,0).
\eea
 As a consequence, we can rewrite the propagator of $\rho$ meson in Eq.(\ref{rhopropagator}) as
 \bea
D^{\mu\nu}_{ab}(q^2) &=&[D_1(q^2)P_{1}^{\mu\nu}+D_2(q^2)P_{2}^{\mu\nu} +D_3(q^2)L^{\mu\nu}+D_4(q^2)u^{\mu}u^{\nu}]\delta_{ab},
\eea
where
\be
D_i(q^2)=\frac{2G_V}{1+2G_V {\Pi}_{i}^2}.
\ee
At last, we use the following gap equations to determine the masses of $\rho$ meson with different spin projections
\be{\label{Gap}}
1+2G_V {\Pi}_{i}^2=0.
\ee

In Eq.(\ref{Lambda}),  the isospin Pauli matrices are $\tau^a=\tau^{\pm}$ and $\tau^b=\tau^{\mp}$ for charged $\rho^{\pm}$ meson, and $\mathbf{\tau}^{\pm}=\frac{1}{\sqrt{2}}(\mathbf{\tau}^{1}\pm i \mathbf{\tau}^{2})$. In the rest frame of $\rho$ meson, i.e., $q_{\mu}=(M_{\rho^{\pm}},\textbf{0})$, the one quark loop polarization function takes the form of
\begin{eqnarray}
\Pi^{\mu\nu}_{\rho^{\pm}}(q^2)&=&i\sum_{p,k=0}^{\infty}\int \frac{dk_{0}dk_{3}}{(2\pi)^3}\int dk_2 dx_1 \Lambda^{\mu\nu}_{\rho^{\pm},{pk}}(\bar{p},\bar{k},x_1)\big{|}_{b.c}.
\end{eqnarray}
After  calculating the $\Pi^{\mu\nu}_{\rho^{\pm}}(q^2)$ in the rest frame of $\rho$ meson (the details are given in \cite{Liu:2014uwa}), we can get the matrix as following:

\bea
\label{eq:matrixrhopm}
\Pi^{\mu\nu}_{\rho^{\pm}}=\left(\begin{matrix}0&0&0&0\cr0&\Pi^{11}&\Pi^{12}&0 \cr0& \Pi^{21}&\Pi^{22}&0\cr0&0&0&\Pi^{33} \end{matrix}\right)
=\left(\begin{matrix}0&0&0&0\cr0&a& ib &0 \cr0& -ib&a&0\cr0&0&0&c \end{matrix}\right).
\eea
Here, we define  $\Pi^{11}=\Pi^{22}=a$, $\Pi^{12}=-\Pi^{21}=ib$ and $\Pi^{33}=c$. 
Note that the calculation in \cite{Liu:2014uwa} was finished in the vacuum. However, in this paper, we are going to investigate the charged $\rho$ condensation at finite temperature and density. Therefore, $k_0$ should be replaced by the Matsubara frequency summation, which is given in the Appendix \ref{Integrals k0}. The matrix elements $\Pi^{11}=\Pi^{22}$, $\Pi^{12}=-\Pi^{21}$ and $\Pi^{33}$ for $\rho^+$ at finite temperature are given as following:
\bea
\label{eq:matrixelement}
&&\Pi^{11}=\Pi^{22}=24\sum_{p,k=0}^{\infty}\int \frac{dk_3}{(2\pi)^2}\int dk_2dx_1\{\mathcal{A}A^+\alpha^++\mathcal{B}A^-\alpha^-\},\nonumber\\
&&\Pi^{12}=-\Pi^{21}=24\sum_{p,k=0}^{\infty}\int \frac{dk_3}{(2\pi)^2}\int dk_2dx_1\{\mathcal{B}(-is_uA^-\alpha^+)+\mathcal{A}(-is_uA^+\alpha^-)\},\nonumber\\
&&\Pi^{33}=24\sum_{p,k=0}^{\infty}\int \frac{dk_3}{(2\pi)^2}\int dk_2dx_1\{\mathcal{D}B^+\beta^++\mathcal{C}B^-\beta^-\}.
\eea
The definitions of $\mathcal{A}$, $\mathcal{B}$, $\mathcal{C}$ and $\mathcal{D}$ are in the Appendix \ref{Integrals k0}, and $A^{\pm}$, $\alpha^{\pm}$, $B^{\pm}$ and $\beta^{\pm}$ are represented as follows:
\beqn
&&A^{\pm}=\frac{1}{2}\left[f_p^{+s_{u}}(x_1)f_k^{+s_{d}}(x_1)\pm\Pi_{p}\Pi_{k}f_p^{-s_{u}}(x_1)f_k^{-s_{d}}(x_1)\right],\nonumber\\
&&\alpha^{\pm}=\frac{1}{2}\left[f_p^{+s_{u}}(0)f_k^{+s_{d}}(0)\pm\Pi_{p}\Pi_{k}f_p^{-s_{u}}(0)f_k^{-s_{d}}(0)\right],\nonumber\\
&&B^{\pm}=\frac{1}{2}[\Pi_{p}f_k^{+s_d}(x_1)f_p^{-s_u}(x_1)\pm\Pi_{k}f_p^{+s_u}(x_1)f_k^{-s_d}(x_1)],\nonumber\\
&&\beta^{\pm}=\frac{1}{2}[\Pi_{p}f_k^{+s_d}(0)f_p^{-s_u}(0)\pm\Pi_{k}f_p^{+s_u}(0)f_k^{-s_d}(0)].
\eeqn
where $\{s_u,s_d\}=\{\text{sgn}(q_u eB),\text{sgn}(q_d eB)\}$ and $\{q_u,q_d\}=\{2/3,-1/3\}$.

Combining the expression in Eq.(\ref{tensordecompose}) and the matrix in Eq.(\ref{eq:matrixrhopm}), we can easily find the relation for charged $\rho^{\pm}$ meson
\bea
\label{Pi^2}
&&{\Pi}^2_{1}=-(a+b),  \nonumber\\
&&{\Pi}^2_{2}=b-a, \nonumber\\
&&{\Pi}^2_{3}=-c.
\eea
As it has been discussed in \cite{Das:1994vr},  $\Pi_4^2$ should be zero in the rest frame of $\rho$ mesons, guaranteed by the Ward identity.

\section{Numerical Results and Discussions}

\subsection{Parameters}

For numerical calculations, we use the soft cut-off functions in Refs.\cite{Liu:2014uwa, Frasca:2011zn}
\bea
&&f_\Lambda=\sqrt{\frac{\Lambda^{10}}{\Lambda^{10}+\mathbf{k}^{2*5}}}, \\
&&f_{\Lambda,eB}^k=\sqrt{\frac{\Lambda^{10}}{\Lambda^{10}+(k_3^2+2|Q eB|k)^5}},
\eea
for zero and nonzero magnetic fields, respectively. At finite magnetic field, we sum up to 20 Landau levels, and the results are saturated. Following Refs.\cite{He:1997gn, Liu:2014uwa}, the parameters of our model, namely the coupling constants $G_S$ and $G_V$, the current quark mass $m_0$, and the three-momentum cut-off $\Lambda$, are determined by reproducing the pion decay constant $f_{\pi}$, the quark mass $M$, the mass of $\pi$, and the mass of $\rho$ in the vacuum. We obtain $\Lambda=582$ MeV, $G_S\Lambda^2=2.388$, $G_V\Lambda^2=1.73$, and $m_0=5$  MeV by choosing $f_{\pi}=95$  MeV, $m_{\pi}=140$ MeV, $M_{\rho}=768$ MeV, the vacuum quark mass $M=458$ MeV.

\subsection{Numerical results for charged $\rho^{\pm}$ and discussions}

In our numerical calculations, we firstly solve the gap equation Eq.(\ref{eq:gapequationmq}) to obtain the quark constitute mass $M$, and then solve the gap equations of Eq.(\ref{Gap}) for charged vector mesons.

\begin{figure}[h!]
\centerline{\includegraphics[width=9cm,height=6cm]{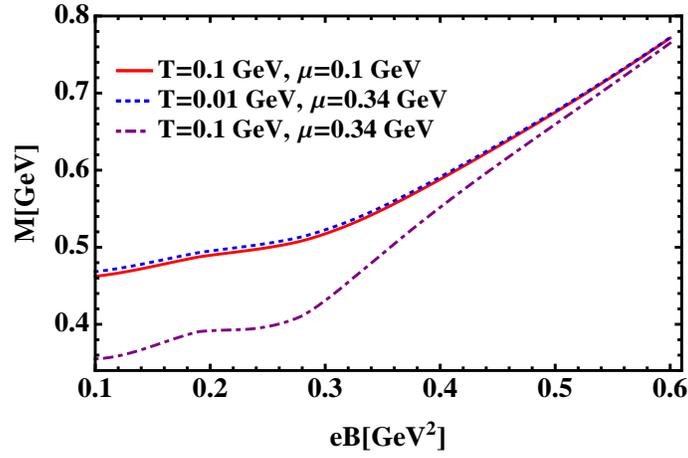}}
\caption{Quark constitute mass $M$ as a function of magnetic field $eB$ with different temperatures T and chemical potentials $\mu$.}
\label{fig:quarkmass}
\end{figure}

\begin{figure}[h!]
\centerline{\includegraphics[width=9cm,height=6cm]{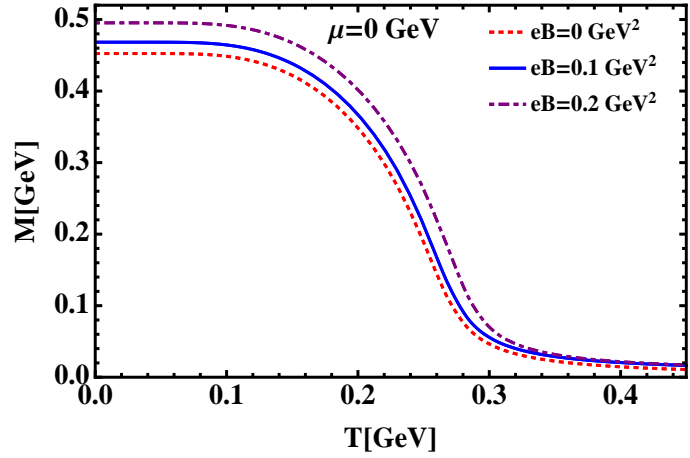}}
\caption{Quark constitute mass $M$ as a function of  the temperature T with different magnetic field $eB$ and zero chemical potential $\mu=0$.}
\label{fig:quarkmassT}
\end{figure}

Fig.~\ref{fig:quarkmass} shows the quark mass dependence of magnetic field $eB$ at several fixed temperatures $T$ and chemical potentials $\mu$,  and Fig.~\ref{fig:quarkmassT} describes the quark mass dependence of the temperature $T$ at $\mu=0$ for several different magnetic fields $eB$. It is noticed  that in the current model, there is no mechanism for inverse magnetic catalysis around the critical temperature. Therefore, Fig.~\ref{fig:quarkmass} and Fig.~\ref{fig:quarkmassT} only show magnetic catalysis behavior of the quark mass.


In Ref.\cite{Liu:2014uwa}, we have observed that there is possible condensation of charged $\rho^+(\rho^-)$ with $s_z=+1(s_z=-1)$ in the vacuum when the magnetic field is stronger than the critical magnetic field $eB_c\simeq0.2$ GeV$^2$. The main goal of this work is to investigate the melting of the charged $\rho$ condensation at finite temperature and chemical potential.

\begin{figure}[h!]
\centerline{\includegraphics[width=9cm,height=6cm]{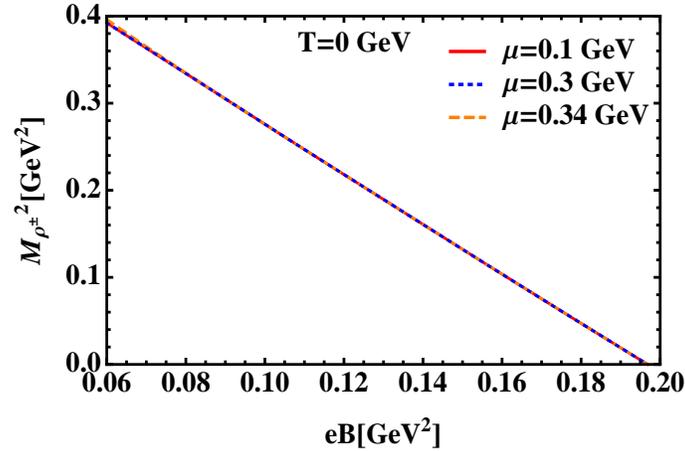}}
\caption{The mass of charged $\rho^+(\rho^-)$ with $s_z=+1(s_z=-1)$ in magnetic field with fixed $T=0$ MeV with different chemical potentials $\mu$.}
\label{fig:rhomassT0}
\end{figure}

Fig.~\ref{fig:rhomassT0} shows the mass of charged $\rho^{\pm}$ as the function of the magnetic field $eB$ at zero temperature for different chemical potentials. It is observed from Fig.~\ref{fig:rhomassT0} that at zero temperature, when $\mu$ is smaller than the vacuum constituent quark mass $M_q^0=458 {\rm MeV}$, the behavior of magnetic field dependence of the charged $\rho^{\pm}$ mass almost does not change for different chemical potentials, and the critical magnetic field remains as $eB_c=0.2 {\rm GeV}^2$. This indicates that the charged $\rho$ condensation can survive at high baryon density and might occur inside compact stars.

\begin{figure}[h!]
\centerline{\includegraphics[width=9cm,height=6cm]{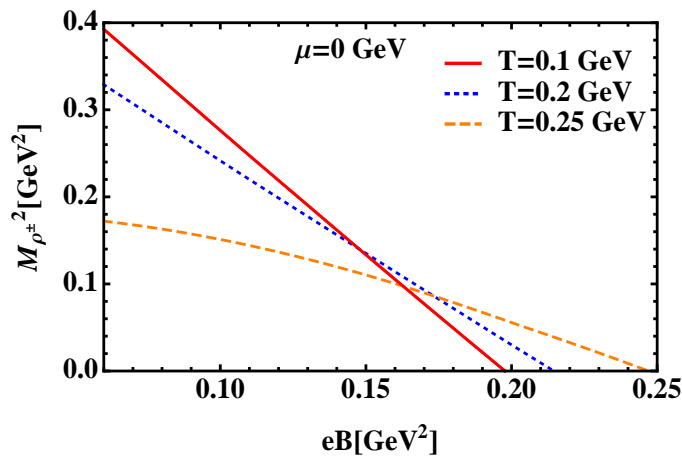}}
\caption{The mass of charged $\rho^+(\rho^-)$ with $s_z=+1(s_z=-1)$ in magnetic field with fixed $\mu=0$ MeV and different temperatures $T$.}
\label{fig:rhomassmu0}
\end{figure}

Fig.~\ref{fig:rhomassmu0} shows the mass of charged $\rho^{\pm}$ as the function of the magnetic field $eB$ at zero chemical potential for different temperatures. It is seen that for different temperatures, the charged $\rho^{\pm}$ mass decreases with the magnetic field and drops to zero at the critical
magnetic field. With the increase of temperature, the critical magnetic field also increases. This indicates that the charged $\rho$ condensation can survive
even at high temperatures.

\begin{figure}[h!]
\centerline{\includegraphics[width=9cm,height=6cm]{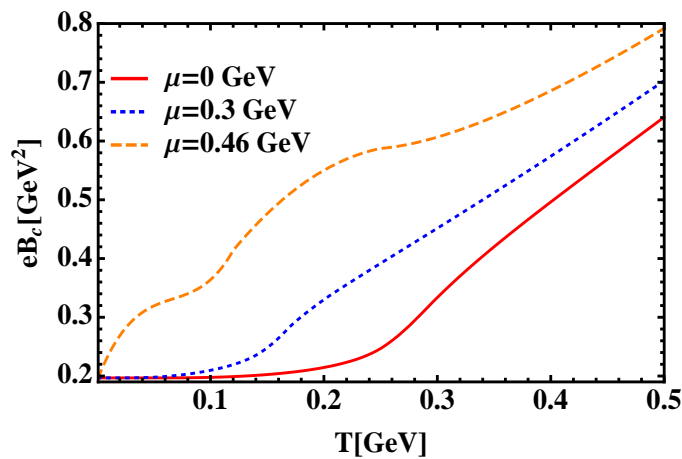}}
\caption{The critical magnetic field $eB_c$ as a function of temperature $T$ with different chemical potential $\mu$ .}
\label{fig:eBcT}
\end{figure}

At last, in Fig.~\ref{fig:eBcT}, we show the critical magnetic field $eB_c$ as a function of the temperature for $\mu=0, 0.3, 0.46 {\rm GeV}$, respectively.
It can be read that when the temperature increases, the critical magnetic field increases. Furthermore, it is seen that the critical magnetic field increases with the chemical potential at fixed temperature. It is worthy of mentioning that at zero chemical potential, when the temperature is below $T=250 {\rm MeV}$, which is almost the critical temperature for the chiral phase transition, we can read that the critical magnetic field does not change so much comparing with its value at zero temperature. However, when the temperature is higher than $T=250 {\rm MeV}$, it is found that the critical magnetic field increases linearly with the temperature, and in the temperature range $200-500~ {\rm MeV}$, the critical magnetic field for charged $\rho$ condensation is in the range of $0.2-0.6~ {\rm GeV}^2$, which is just located inside the range of the magnetic field generated in the non-central heavy ion collisions of LHC. It means that high temperature superconductor could be produced at LHC.

\section{Conclusions}

We have investigated the charged $\rho$ mesons in an external magnetic field at finite temperature and density by using the NJL model. The mesons are constructed by summing up infinite quark-loop chains by using the random phase approximation. In this paper, we calculate the $\rho$ meson polarization tensor to the leading order of  $1/N_c$ expansion, i.e.  the one quark loop. In this process, the constituent quark mass is solved self-consistently with magnetic field at finite temperature and density. It is noticed that in our current framework, there is no inverse magnetic catalysis mechanism, so the quark mass increases with magnetic field and only shows the magnetic catalysis effect.

The mass of charged $\rho$ meson depending on the magnetic field is calculated at finite temperature and density. It is found that at fixed temperature and density, the polarized charged $\rho$ meson becomes lighter with the increase of magnetic fields and drops to massless at the critical magnetic field $eB_c$, which indicates that there will appear charged $\rho$ meson condensation when the strength of  the magnetic field is greater than the $eB_c$. Moreover, it is observed that the critical magnetic field $eB_c$ increases with both the temperature and density. Particularly, at zero density, our results show that in the  temperature region $200~{\rm MeV}<T<500~ {\rm MeV}$, the critical magnetic field $eB_c$ is in the range of $0.2-0.6~ {\rm GeV}^2$, which indicates that the high temperature superconductor could be created in the early stage of LHC.

However, we have to mention that in our current framework, there is no inverse magnetic catalysis for the quark mass. In the next step, we will investigate how the inverse magnetic catalysis \cite{IMC} will affect our results on the charge $\rho$ condensation, especially at high temperatures.

\section{Acknowledgement}
We thank J.F. Liao and I. Shovkovy for valuable discussions. This work is supported by the NSFC under Grant Nos. 11275213, and 11261130311(CRC 110 by DFG and NSFC), CAS key project KJCX2-EW-N01, and Youth Innovation Promotion Association of CAS. L.Yu is partially supported by China
Postdoctoral Science Foundation under Grant No. 2014M550841.

\appendix

\section{Integrals of $k_0$}
\label{Integrals k0}
We introduce the notations of the integrals of $k_0$ for $\rho^+$ meson as follows:
\bea
\label{intk_0}
&&\mathcal{A}= i\int\frac{dk_0}{2\pi}\frac{\bar{p}_{(u)} \cdot\bar{k}_{(d)}-M^2}{(p_0^2-\omega_{u,p}^2)(k_0^2-\omega_{d,k}^2)} =\frac{1}{2}(iI_1+iI_1^{'})\nonumber\\&& -\left[\frac{1}{2}\left(M_{\rho^+}^2-{\bar{p}_{(2,u)}}^2
-{\bar{k}_{(2,d)}}^2\right) +\bar{p}_{(2,u)}\bar{k}_{(2,d)}\right]iI_2, \nonumber\\
\eea
\bea
&&\mathcal{B}= i\int\frac{dk_0}{2\pi}\frac{\bar{p}_{(u)}\cdot\bar{k}_{(d)}-M^2+2\bar{p}_{(2,u)}\bar{k}_{(2,d)}}
{(p_0^2-\omega_{u,p}^2)(k_0^2-\omega_{d,k}^2)}
 =\frac{1}{2}(iI_1+iI_1^{'})\nonumber\\&&-\left[\frac{1}{2}\left(M_{\rho^+}^2-{\bar{p}_{(2,u)}}^2
-{\bar{k}_{(2,d)}}^2\right)-\bar{p}_{(2,u)}\bar{k}_{(2,d)}\right]iI_2, \nonumber \\
\eea
\bea
&&\mathcal{C}= i\int\frac{dk_0}{2\pi}\frac{p_0k_0+\bar{p}_{(2,u)}\bar{k}_{(2,d)}+k_3^2-M^2}{(p_0^2-\omega_{u,p}^2)(k_0^2-\omega_{d,k}^2)}
=\frac{1}{2}(iI_1+iI_1^{'})\nonumber\\&&+\left[2k_3^2-\frac{1}{2}\left(M_{\rho^+}^2-{\bar{p}_{(2,u)}}^2-{\bar{k}_{(2,d)}}^2\right)+\bar{p}_{(2,u)}\bar{k}_{(2,d)}\right]iI_2,\nonumber\\
\eea
\bea
&&\mathcal{D}=i\int\frac{dk_0}{2\pi}\frac{p_0k_0-\bar{p}_{(2,u)}\bar{k}_{(2,d)}+k_3^2-M^2}{(p_0^2-\omega_{u,p}^2)(k_0^2-\omega_{d,k}^2)} =\frac{1}{2}(iI_1+iI_1^{'})\nonumber\\&&+\left[2k_3^2-\frac{1}{2}\left(M_{\rho^+}^2-{\bar{p}_{(2,u)}}^2-{\bar{k}_{(2,d)}}^2\right)
-\bar{p}_{(2,u)}\bar{k}_{(2,d)}\right]iI_2.\nonumber\\
\eea
Here, $\omega_{u,p}^2=2|q_u eB|p+p_3^2+M^2$, 
   $\bar{p}_{(2,u)}=-s_{u}\sqrt{2|q_{u}eB|p}$ 
and  $\bar{p}_{(u)}=(p_0,0,-s_{u}\sqrt{2|q_{u}eB|p},p_3)$. 
We use the similar definition for  
  $ \bar{k}_{(d)}$, 
   $ \bar{k}_{(2,d)}$, $\omega_{d,k}$.
Moreover, $I_1$, $I_1^{'}$ and $I_2$ are given by:
\bea
&&I_1=\int\frac{dk_0}{2\pi}\frac{1}{k_0^2-\omega_{u,p}^2}, \\
&&I_1^{'}=\int\frac{dk_0}{2\pi}\frac{1}{k_0^2-\omega_{d,k}^2}, \\
&&I_2=\int\frac{dk_0}{2\pi}\frac{1}{((k_0+q_0)^2-\omega_{u,p}^2)(k_0^2-\omega_{d,k}^2)}.
\eea

As in Ref.\cite{Rehberg:1995nr}, we replace the integration over $k_0$ by Matsubara summation according to the prescription $\int \frac{dk_0}{2\pi}(\cdots)~ \overrightarrow{} ~iT \sum_{m=-\infty}^{\infty}(\cdots)$ and obtain
\bea
&&iI_1=-(\frac{n_f(\omega_{u,p}-\mu)+n_f(\omega_{u,p}+\mu)-1}{2\omega_{u,p}}),\\
&&iI_1^{'}=-(\frac{n_f(\omega_{d,k}-\mu)+n_f(\omega_{d,k}+\mu)-1}{2\omega_{d,k}}),
\eea

\bea
iI_2&=&-\left[\frac{n_f(\omega_{d,k}-\mu)}{2\omega_{d,k}}\frac{1}{(\omega_{d,k}+M_{\rho^+})^2-\omega^2_{u,p}}\right.\nonumber\\&&\left.-\frac{n_f(-\omega_{d,k}-\mu)}{2\omega_{d,k}}\frac{1}{(-\omega_{d,k}+M_{\rho^+})^2-\omega^2_{u,p}}\right.\nonumber\\
&&\left.+\frac{n_f(\omega_{u,p}-
\mu)}{2\omega_{u,p}}\frac{1}{(-M_{\rho^+}+\omega_{u,p})^2-\omega^2_{d,k}}\right.\nonumber\\&&\left.-\frac{n_f(-\omega_{u,p}-\mu)}{2\omega_{u,p}}\frac{1}{(-M_{\rho^+}-\omega_{u,p})^2-\omega^2_{d,k}}\right],\nonumber\\
\eea
with
\bea
n_f(x)=\frac{1}{1+e^{\frac{x}{T}}}.
\eea
Here, the notations are only for $\rho^+$ meson, and the similar ones can be used for $\rho^-$ meson.

\end{document}